\documentstyle[prl,aps,multicol]{revtex}

\def\be{\begin{equation}}
\def\ee{\end{equation}}

\begin{document}
\draft

\title{Localization in non-chiral network models for two-dimensional
disordered wave mechanical systems} 
\author{Peter Freche, Martin
Janssen, Rainer Merkt \\\ Institut f\"ur Theoretische Physik,
Universit\"at zu K\"oln, \\\ Z\"ulpicher Strasse 77, 50937 K\"oln,
Germany } 
\date{August 6 1998} \maketitle

\begin{abstract} 
Scattering theoretical network models for general coherent wave
mechanical systems with quenched disorder are investigated.  We focus
on universality classes for two dimensional systems with no preferred
orientation: Systems of spinless waves undergoing scattering events
with broken or unbroken time reversal symmetry and systems of spin 1/2
waves with time reversal symmetric scattering.  The phase diagram in
the parameter space of scattering strengths is determined.  The model
breaking time reversal symmetry contains the critical point of quantum
Hall systems but, like the model with unbroken time reversal symmetry,
only one attractive fixed point, namely that of strong localization.
Multifractal exponents and quasi-one-dimensional localization lengths
are calculated numerically and found to be related by conformal
invariance. Furthermore, they agree quantitatively with theoretical
predictions.  For non-vanishing spin scattering strength the spin 1/2
systems show localization-delocalization transitions.
\end{abstract}


\begin{multicols}{2}

Large intensity fluctuations and spatial localization
are fascinating universal features in any coherent wave mechanical system
subject to quenched disorder.  A modeling that covers essential
symmetries and characteristic length scales, but does not 
rely on
particular dispersion relations and specific details is highly
desirable in various fields of theoretical physics, e.g.
in optics, mesoscopic electronics, and quantum chaos (see e.g.
\cite{IshBeeCas}). Already in 1982 Shapiro \cite{Sha82} pointed out that
a convenient modeling of disordered coherent wave mechanical systems 
can be given by networks of unitary random scattering matrices. Only
recently it was fully appreciated 
 that such network models (NWMs) have a number
of advantages over more traditional Hamiltonian models (e.g. the
Anderson model used in mesoscopic electronics or the Helmholtz equation
with random refraction index used in optics). NWMs
 yield directly transport quantities \cite{Met98},
 wave packet dynamics \cite{Huc98}, and quasi-energy eigenvalues and
 eigenstates \cite{Kle97}.  Furthermore, real space renormalization
 group (RG) treatments of NWMs \cite{Gal97} open new
 perspectives for investigating localization-delocalization (LD)
 transitions.  A well known NWM introduced by Chalker and Coddington (CC-model)
 \cite{Cha88} (see also \cite{Fer88}) describes the situation of disordered
 two-dimensional  (2D) electrons undergoing the quantum Hall LD transition
 in a strong perpendicular magnetic field.  

In contrast to the CC-model
 here we deal with systems in the absence of  fields that would introduce a
handedness (chirality). However we do allow for  the
 breaking of time reversal symmetry. 
 In particular, we address three classes of non-chiral (NC)
 2D NWMs with respect to their quantum localization
 properties. The first model describes time reversal symmetric
scattering and is denoted as
 O2NC-model, where 'O' stands for 'orthogonal' (a corresponding
 Hamiltonian can be diagonalized by orthogonal matrices).  Second, a
 similar model with broken time reversal symmetry  ('U' for
'unitary') is introduced denoted as
 U2NC-model. It describes e.g. the motion of
mesoscopic (spinless) electrons in the presence of random magnetic
fields with zero mean and disorder potentials. The third  model deals with
  time reversal
 symmetric scattering also in spin degrees of freedom. It is denoted as
 S2NC-model ('S'  for 'symplectic') and a detailed
 analysis for this model is given in \cite{Mer98}.  Here we focus
 on the U/O2NC-model, discuss their construction and  the phase diagram in
 detail, and present  a quantitative check of analytical results
 for quasi-1D localization lengths, multifractal exponents and
 conformal invariance  in the weak localization
 regime \cite{Note1}.

Quite generally, a NWM can be constructed as follows. Take a regular
network of ${\cal N}$ sites and $N$ bonds.  Each bond $\alpha$ carries
propagating wave modes ($n^{+}_\alpha$ incoming modes and
$n^{-}_\alpha$ outgoing modes) represented by complex amplitudes,
$\psi_{n^{\pm}_\alpha}$.  On the sites unitary $S$-matrices map
incoming to outgoing amplitudes.  The elements of each $S$-matrix are
(in general) random quantities respecting the symmetries of $S$ and are
characterized by typical scattering strengths (see Fig.~1).  Random phases are
attached to the amplitudes on the bonds. They simulate the random
distances between scatterers in realistic systems.

The construction of a NWM is fixed by the choice of a certain type of
random $S$-matrix and a ${\cal N}\times {\cal N}$ connectivity matrix
$C$ which has elements $C_{ij}=1$ if a wave mode can propagate from
site $i$ to site
$j$  and $C_{ij}=0$ otherwise.  The NWM defines a
dynamical system since  $S$ and $C$ give rise to  a unitary
matrix $U$ \cite{Fer88,Edr88} that maps all incoming to outgoing 
wave modes. Its dimension $B$ equals the total number of amplitudes
defined on  bonds,
$B\propto N$. We choose a convenient time unit and  denote
the vector of the $B$  amplitudes as $\psi$ such that 
 $\psi(t+1)=U \psi(t)$.  The eigenphases
$\phi_n$ ($n=1\ldots B$) and corresponding eigenvectors $\psi_n$ of
$U$, $ U\psi_n = \exp(i\phi_n) \psi_n \, , $ can then be  interpreted as
quasi-energies and corresponding eigenstates.

Let us now report on the U2NC- and the O2NC-model. The corresponding
$S$-matrix is graphically represented in Fig.~1a. It maps four
ingoing modes to four outgoing modes consistent with the connectivity
matrix that corresponds to a regular   2D
square lattice where each site is connected to four nearest neighbors.
Apart from random phases the parameters of each $S$ matrix are the
strengths of transmission $t$, reflection $r$, and  deflection
$d$ ($r,t$ and $d$ denote the modulus of the corresponding
scattering amplitudes)\cite{Note3}.
Deflection to the left and right are set
equal.  This choice classifies the model to be non-chiral. 
 The unitarity of the $S$-matrix requires
$t^2+r^2+2d^2=1$ and the corresponding parameter space shown in
Fig.~1b forms the phase space of the U2NC-model.  For time reversal
symmetric scattering (O2NC-model) the $S$-matrix is symmetric and the
phase space is reduced  by $r+t\geq 1$.

The change of the total scattering strengths  of a finite
 network  with respect to system
 size yields a RG flow in phase space.  It is evident
 that there are (at least) three fixed points of the RG  flow in
 the U2NC-model, namely the localization fixed point (LFP) with $r=1$, the
 metallic fixed point (MFP) with $t=1$, and the Chalker-Coddington
 fixed point (CC-FP) with $d^2=0.5$.  Furthermore, the system consists
 of uncoupled one-dimensional chains for $d=0$. The  LFP corresponds to
 complete localization and the MFP corresponds to  perfectly extended
 states. That $d^2=0.5$ corresponds to a fixed point follows from the
 observation that the network at $t=r=0$ is, in fact, a decoupled set
 of two uncorrelated CC-models right at their critical  points.

  We investigated the LD properties by the transfer matrix technique
(similar as in \cite{Cha88}) and calculated localization lengths (LLs)
$\xi(M)$ for quasi-1D strip geometries of width $M$. We took system
lengths in the range from $10^5$ to $10^6$ leading to statistical
errors in LLs of about $0.5\%$ -- $5\%$.  Dividing by $M$ yields a
convenient scaling variable $ \Lambda(M,r,t)=\xi(M,r,t)/{M}$. 
For $\Lambda$ increasing (decreasing) with $M$ the 2D system will
be in a delocalized (localized) phase; at the critical point of a LD
transition $\Lambda$ is independent of $M$. Note that  2D
systems with large conductance 
always behave  similar to critical systems since $\Lambda$ is
almost constant. However, in contrast to generic critical points,
$\Lambda$ is large and shows ``weak localization'' (or ``weak
anti-localization'') effects.  An analytic expression, valid for
$\Lambda\gg 1$, can be obtained from a Fokker-Planck approach to
disordered quasi-1D systems. It is characterized by the number
$N_c\propto M$ of transverse quantum channels (each of which splits
into a right moving and a left moving wave mode) and an elastic mean
free path $l_e$.  It reads \cite{Mac92} $ \xi(M)=l_e(\beta N_c +2
-\beta) $ where $\beta= 1,2,4$ for orthogonal, unitary and 
symplectic symmetry, respectively.  The main effect of
symmetry is reflected by the first contribution, $\Lambda_0=\beta
l_e(N_c/M)$, to $\Lambda$.  The second contribution, $l_e(2-\beta)/M$,
describes the sign of the RG-flow equation $d\Lambda/d(\ln M) = -
(\Lambda -\Lambda_0)$, hence the weak localization (anti-localization)
correction for $\beta=1$ ($\beta=4$). The  localization
correction for $\beta=2$ is not contained in this approximation.

Fig.~2 displays a contour plot of the function $\Lambda(r,t)$ for
the U2NC-model with $M=32$.  $\Lambda$ increases dramatically
for $t\to 1$, decreases for $r\to 1$ and vanishes
at $r=1$. For $r,t\to 0$, it drops down to the value
$\Lambda\approx 1.23 \pm 0.02$ which is known as the critical value in
the CC-model (see e.g. \cite{Doh96}). Varying $M$ leads to
qualitatively similar contour plots, leaving the three fixed point
values constant (up to small finite-size effects). We found no
indication for the existence of further fixed points.  Instead we
found that LLs converge to finite values $\xi_\infty$ as long as
scattering parameters are in a regime close to the LFP. As expected there,
the scaling flow towards the LFP can be described by a one-parameter
scaling function. The parameter is just $\xi_\infty(r,t)/M$. The same
qualitative findings, indicating strong localization in the
thermodynamic limit, were obtained for the restricted phase space of
the O2NC-model (see Fig.~1b).  Close to the CC-FP in the U2NC-model,
where $r,t \ll 1$, a different one-parameter scaling function of the
parameter $M^2(r^2+t^2)$ gives a reasonable fit to the RG flow away from the
CC-FP. This parameter can be interpreted as the total tunneling
probability between two critical CC-models that are uncoupled at
$r=t=0$.

In the regime close to the MFP we found weak localization corrections.
For a quantitative analysis we defined the elastic mean free path
in our model in a way consistent with the notions of the quasi-1D
Fokker-Planck approach, $ l_e/\delta L ={\cal
T}/(1-{\cal T})$, where $\cal T$ denotes the transmission strength of
a microscopic scattering array of extension $\delta L$. For the
diagonal arrangement of the network in the transfer matrix
calculations a suitable  microscopic scattering array is the one shown in
Fig.~1a. It corresponds to a length $\delta L=1/2$ and yields 
\be
	l_e=\frac{1}{2}\frac{t^2+d^2}{r^2+d^2}\label{5} \, .  
\ee 
The scattering array has a Landauer conductance (${\cal T}/(1-{\cal T})$
per channel) that we denote as microscopic conductance, $g_{\rm mi}=
4l_e$. A classical conductance $g_{\rm cl}$ can also be defined by
using the Einstein relation (in atomic units), $g_{\rm cl}=2\pi \rho
D$. Here $\rho$ denotes the density of states per unit volume in
(quasi-)energy space, $D=vl_e/2$ the classical diffusion constant,
and $v$ the wave mode velocity.  By construction we have $v=1$,
$B=4{\cal N}$. The range of quasi-energies is $2\pi$. Thus, we find
for the U/O2NC-NWM $g_{\rm cl}=2l_e= g_{\rm mi}/2$.
The channel number for a given width $M$ is $N_c=2M$. Thus,
in the regime of large $\Lambda$, we expect 
\be 
	\Lambda = g_{\rm mi} \frac{2\beta}{4} + \frac{2-\beta}{4M} 
	\label{6} 
\ee 
to be a good approximation to our data.  
In Fig.~3a we have plotted about $300$ values of $\Lambda$ versus
$g_{\rm mi}$ for the U2NC-model obtained 
for $M=4,8,16,32$ and $r,t$ values distributed over the entire phase
space.  For $\Lambda {_> \atop ^\sim} 3$ almost all data points are
consistent with Eq.~(\ref{6}) (for $\beta=2$).  We checked that those
events which are clearly off the line $\Lambda=g_{\rm mi}$ belong to
very low values of $d$ and thus correspond to almost decoupled 1D
chains.  For low values of $\Lambda$ (see inset of Fig.~3a) strong
localization can be observed.  We made similar observations for the
O2NC-model.

As a second quantitative test we calculated the scaling exponent
$\alpha_0$ describing the scaling of the typical local density of
states, $\rho_{\rm t} \propto L^{2-\alpha_0}$, in a square geometry
for which the 2D LL $\xi$ is much larger than the (linear) system size
$L$.  In 2D these states behave like critical states and are
multifractal (for review see \cite{JanR98}).  Non-linear
$\sigma$-model calculations yield analytic results for multifractal
exponents in this regime, and the result for $\alpha_0-2$ reads
\cite{Fal95} 
\be 
	\alpha_0-2=(\pi \beta g_{\rm cl})^{-1}\, .\label{8}
\ee 
This result can also directly be
obtained from Eq.~(\ref{6}) and a conformal mapping relation between
quasi-1D LLs and multifractal exponents
(successfully tested in 2D quantum Hall systems at criticality
\cite{Doh96}). For $\alpha_0-2$ this relation yields $ \alpha_0-2 =
(\pi \Lambda)^{-1}\, .  $ Together with Eq.~(\ref{6}), in the limit
where $\Lambda$ is approximately independent of $M$, this leads to
Eq.~(\ref{8}) \cite{Note2}.  In Fig.~3b the calculated values of
$\alpha_0$ are plotted versus the values obtained from the conformal
mapping relation, denoted as $\alpha_0^c$. The
relation is fulfilled for  $\alpha_0\simeq 2$, 
i.e. for large values of $\Lambda$. Also the critical point of
the CC-model fulfills the relation within the errors.  As expected, for
intermediate and small values of $\Lambda$ the localization
corrections -- in particular for the O2NC-model -- lead to deviations.

Finally, we report briefly on our findings in the S2NC-model (for
details see \cite{Mer98}).  It is based on the O2NC-model. 
In addition
to the scattering parameters $t$ and $r$ a new parameter, the spin
scattering parameter $s\in [0,1]$, appears. As can be seen in Fig.~4a
each scattering matrix of the O2NC-model has twice  the number  of
incoming and outgoing modes to take care of the spin degree of
freedom. The $S$-matrix denoted as `potential' 
belongs to two superimposed
uncoupled O2NC-models.  A time reversal symmetric spin scattering
process can occur, however, at each $S$-matrix of the `spin'
type. This spin $S$-matrix is characterized by the spin scattering
parameter $s$. For $s=1$ an entering spin mode becomes randomized in
spin space after a single scattering event. The peculiar time reversal
behavior of spin $1/2$ states leads to the effect of
weak anti-localization (see e.g. \cite{Ber84}).  This turns the MFP into an
attractive fixed point, hence the occurrence of a LD transition in 2D
seems to be unavoidable, and was observed
in numerical calculations for Hamiltonian models (e.g. \cite{EvaSch}).
For fixed spin scattering parameter $s$ we determined via the transfer
matrix method the phase diagram in $(r,t)$-space.  For any $s>0$ a
metallic phase appears as soon as $r$ is small and $t$ is large enough
(see Fig.~4b).  The area of the metallic phase gradually shrinks to
zero for $s\to 0$.  On fixing two of the parameters, $t=0.6$ and $s=0.4$, and
varying $r$ we obtained a scaling function (by an iterative fit procedure)
for the LD transition, the quality of which was checked by a
$\chi^2$-test.  The scaling function describes the quantity
$\Lambda(r,M)$ as a two-branch function $f^{\pm}(\xi_c(r)/M)$ where
$\xi_c(r)$ is  the correlation length of the LD
transition.  We confirmed the assumption that the LD transition is
governed by one-parameter scaling.  Approaching the transition point
$r^\ast$ the correlation length diverges, $\xi_c\propto
|r-r^\ast|^{-\nu}$, with critical exponent $\nu=2.51
\pm 0.18$.  With the help of the conformal mapping relation 
$\alpha_0=2.174 \pm 0.003$ follows from the
critical value $\Lambda^\ast=1.83\pm 0.03$ obtained as the branching
point of the scaling function.

In conclusion, we have studied non-chiral network models for coherent
wave mechanical systems with quenched disorder in two dimensions.  In
the parameter space of scattering strengths we found three fixed
points of a finite size scaling flow: a metallic fixed point, a
localization fixed point and the quantum Hall fixed point. The latter
is part of phase space only in systems which break time reversal
symmetry (U2NC-model).  Only the localization fixed point is
attractive and thus almost all states localize in the thermodynamic
limit. Furthermore, we showed that a small amount of time reversal
symmetric spin scattering (S2NC-model) turns the metallic fixed point
into an attractive one, and  
metallic and localized phases occur.

We thank A. Altland, A. Dohmen, B. Huckestein, R. Klesse, M. Metzler and in
particular B. Shapiro for useful discussions.  This work was supported
by the DFG within the SFB 341.

\bigskip
\bigskip

\leftline{\bf FIGURE CAPTIONS:}

\noindent
Figure 1: a) The scattering unit of the U/O2NC-model. The modulus of
scattering amplitudes for transmission is $t$, for reflection $r$ and
 for deflection $d$. b) Phase space of the U/O2NC-models.
 The area in light grey is only
available in the U2NC-model. The dark grey area corresponds to
parameters available in both models. Fixed points  are
marked by bullets and  decoupled 1D chains correspond
to the  line $r^2+t^2=1$.

\noindent
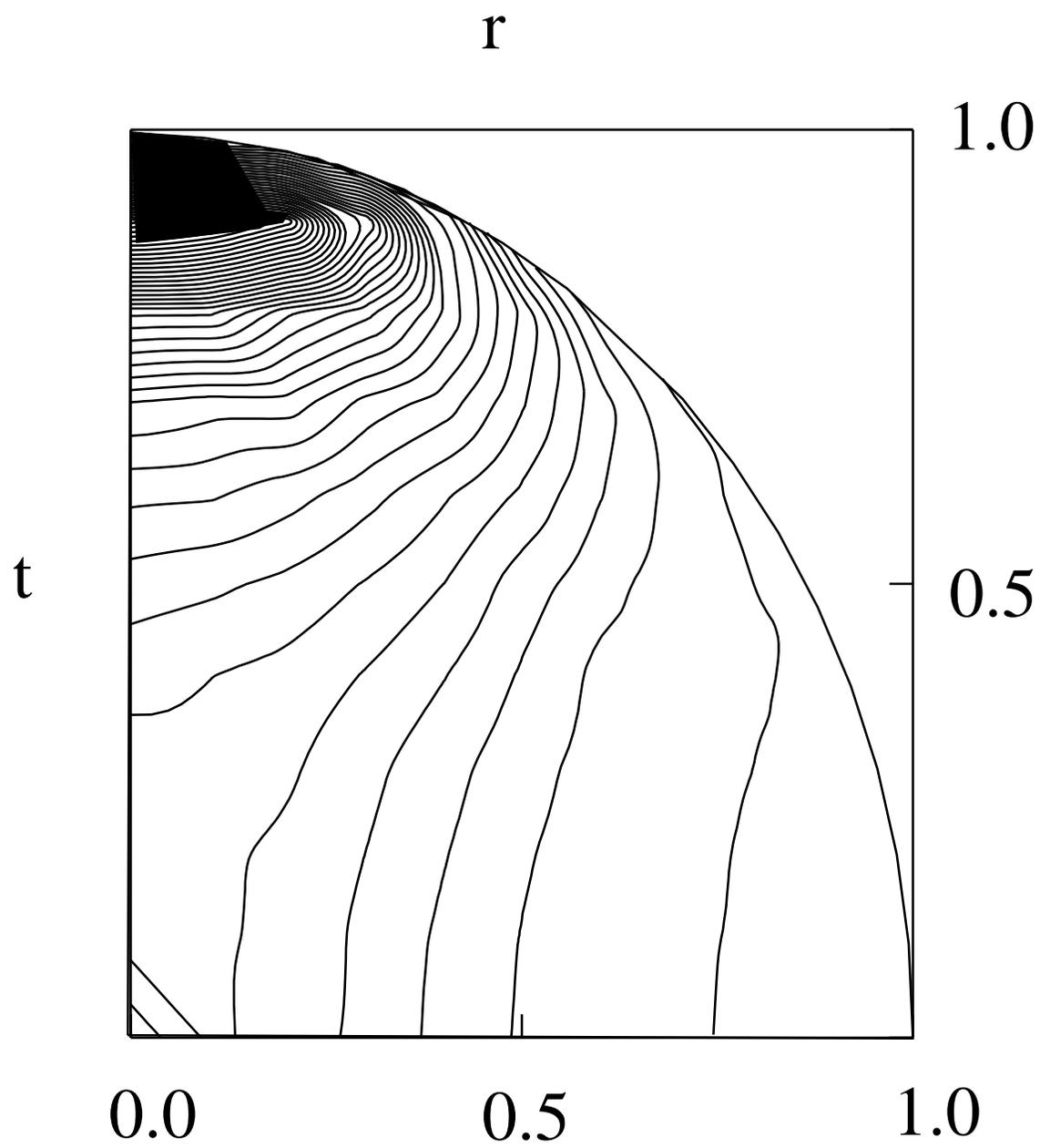
Figure 2: Contour plot of the scaling variable $\Lambda$ in the
parameter space of the U2NC-model for a fixed system width $M=32$.

\noindent
Figure 3: Weak Localization effects: 
a) $\Lambda$  for the U2NC-model as a function of $g_{\rm mi}$
for system widths of $M=4,8,16,32$ and varying values of $r\in
[0,0.9],t\in [0,0.9]$. The inset shows an enlargement for $\Lambda<2$.
b) The numerically obtained
multifractal exponents $\alpha_0$ corresponding to
 different scattering
parameters,   plotted versus 
$\alpha_0^c$ calculated by Eq.~(\ref{8}).

\noindent
Figure 4: S2NC-model: a) Section of  the $S$-matrix for the S2NC-NWM. 
In contrast to  the O2NC-NWM the number of channels 
per bond is doubled. Straight lines correspond to spin-up, dashed
lines to spin-down.  Potential scatterers (light grey) can change the
direction of the  motion of electrons while spin scatterers can only
scatter in the  spin degrees of freedom. 
b) Phase diagram 
of the transition
 for spin scattering strength $s=0.1$. The grey area shows the
delocalized regime and the white area the localized regime. 

\end{multicols}
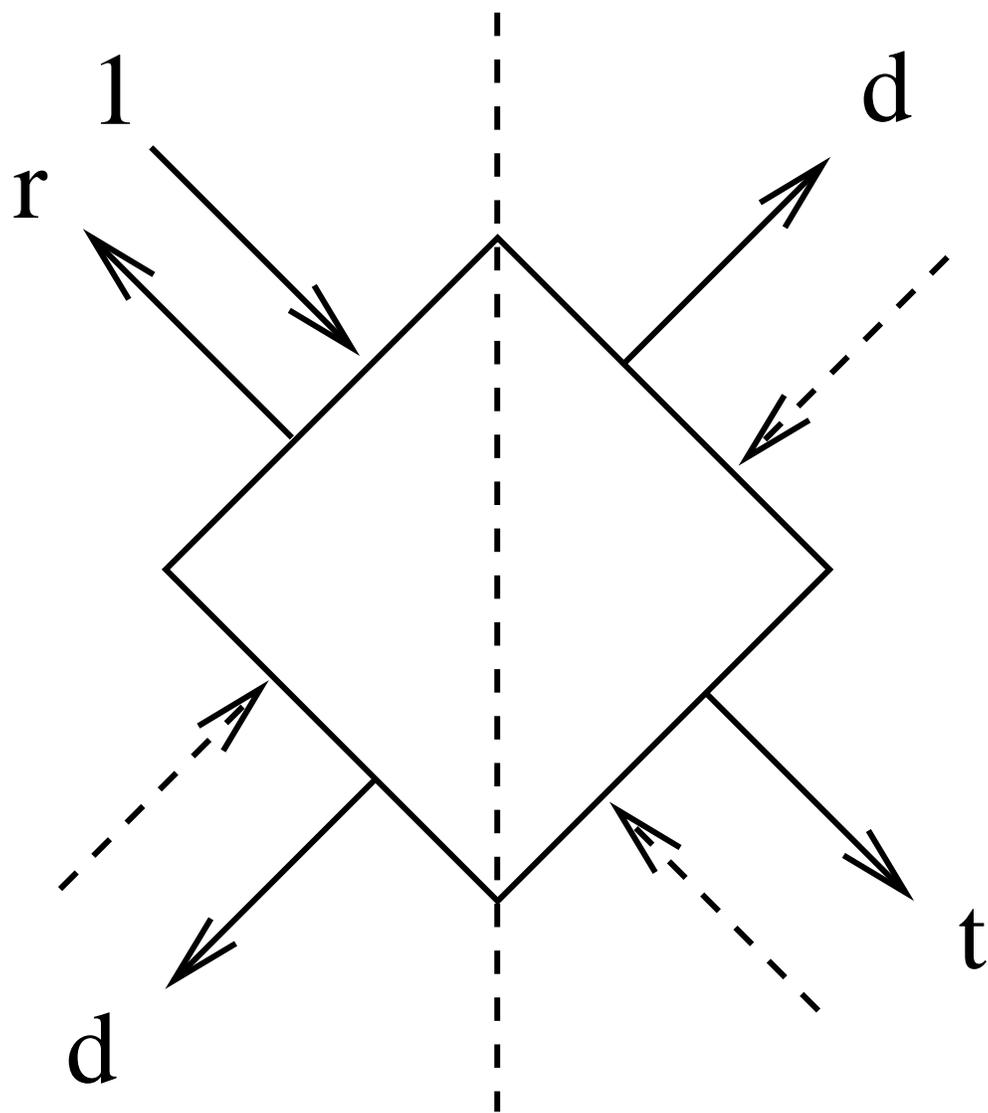
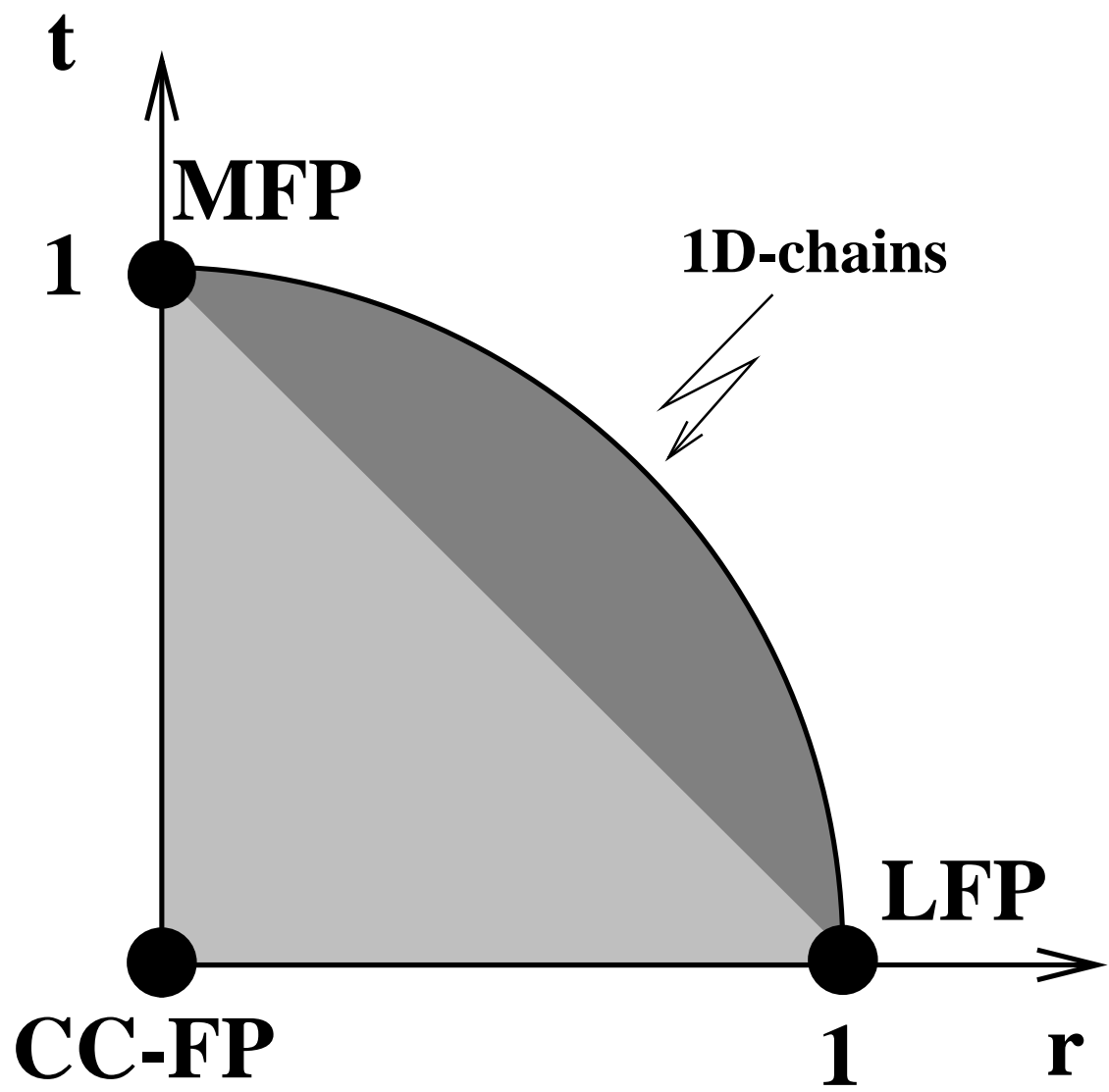
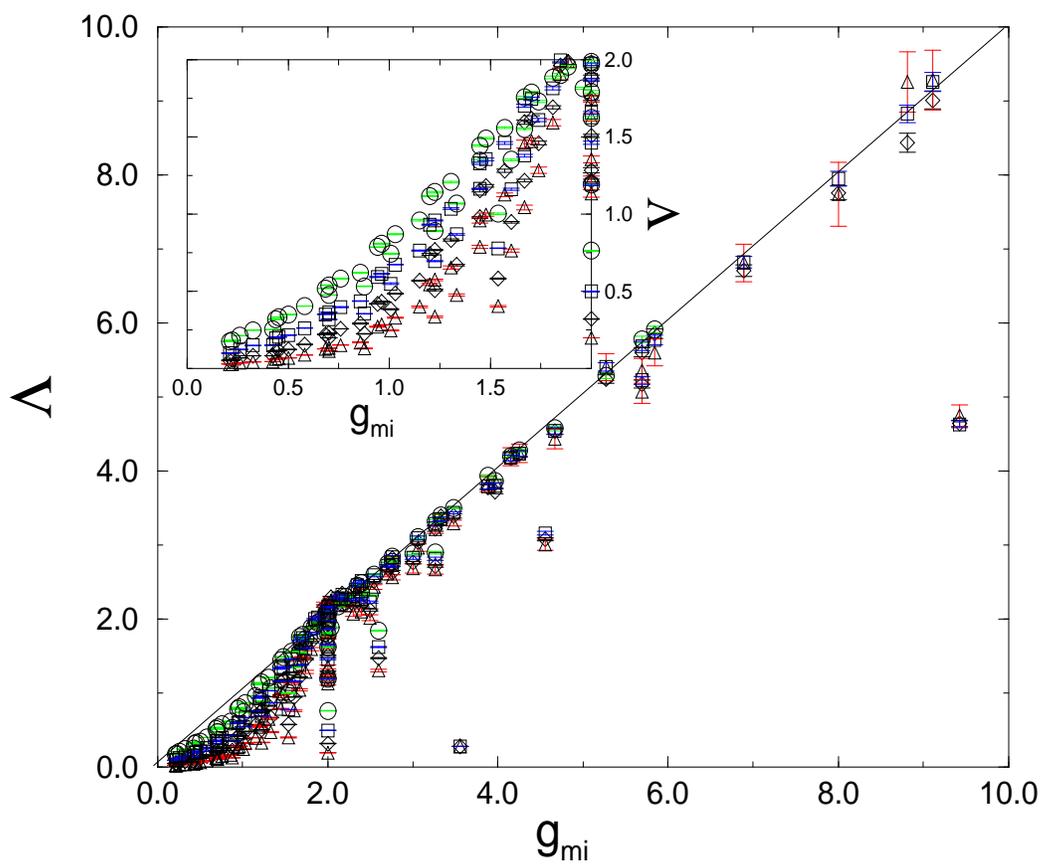
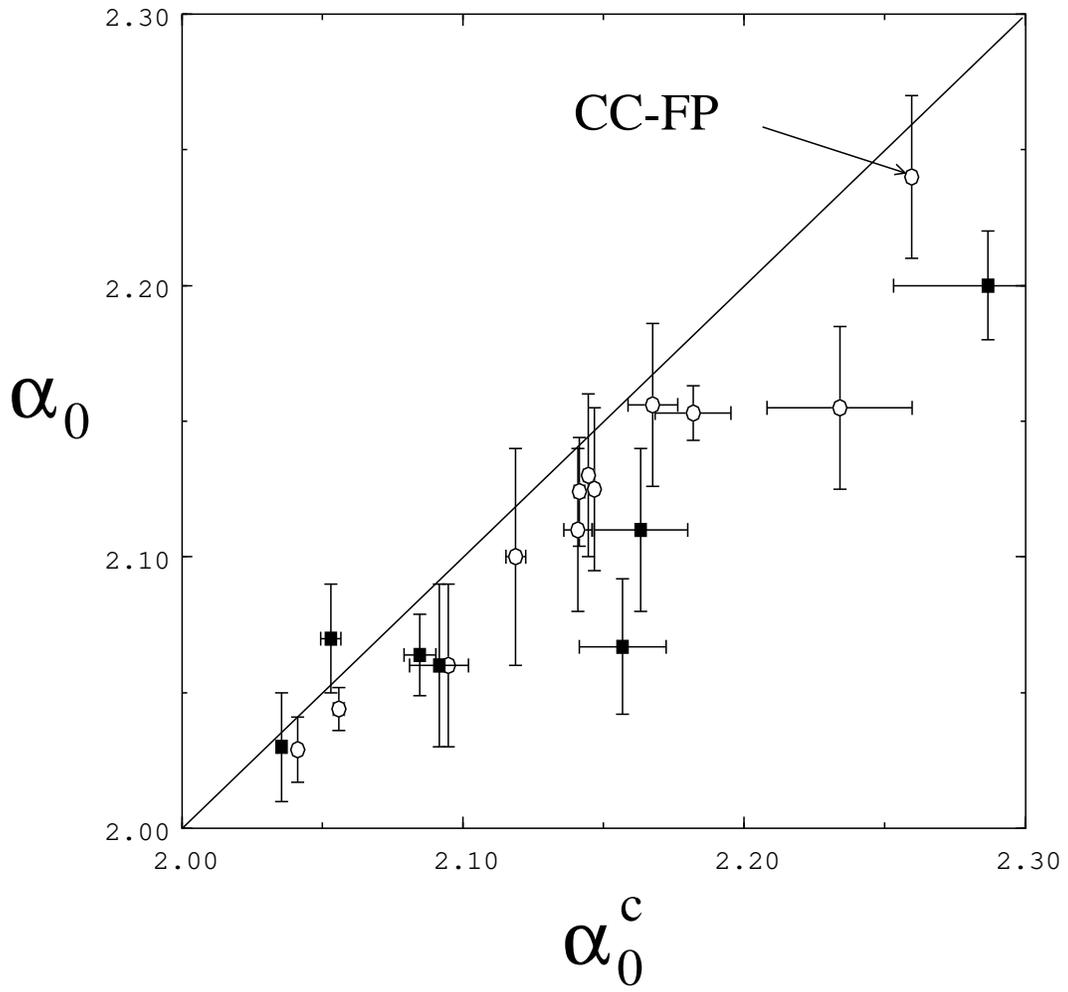
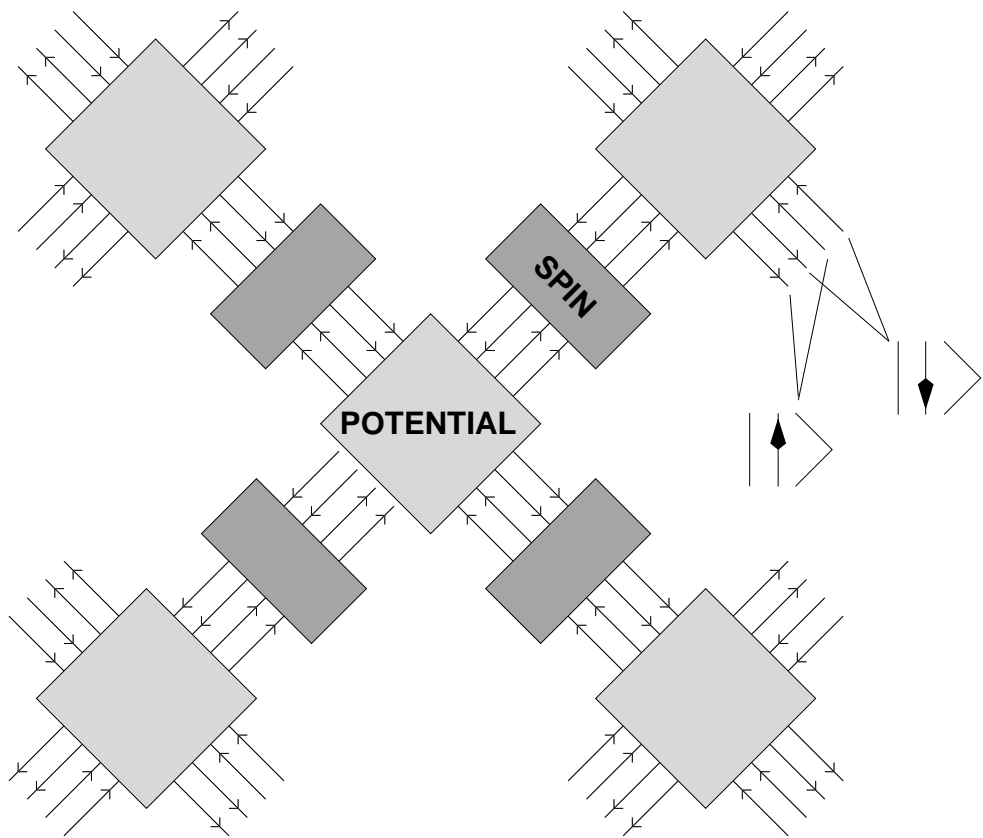
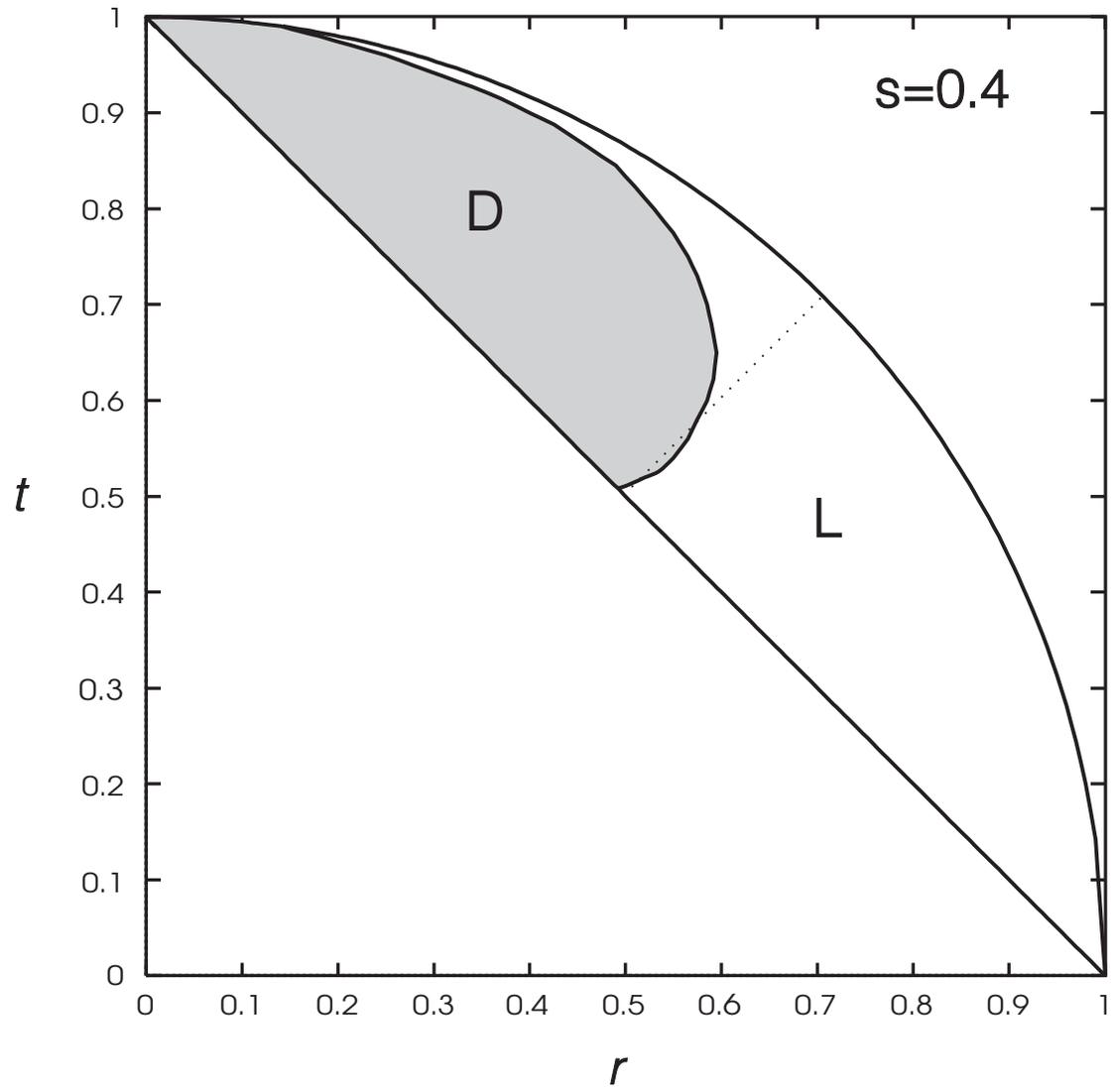
\end{document}